\documentclass[twoside,superscriptaddress,twocolumn]{revtex4}

\usepackage{graphicx, latexsym, amssymb, amsmath, color, multirow,
mathrsfs, CJK, ifpdf}

\usepackage[section]{placeins}
\usepackage{amsmath}
\usepackage{amsfonts}
\usepackage{amssymb}
\usepackage{bm}% bold math
\usepackage{graphicx}
\usepackage{color} %支持彩色字体
\usepackage{txfonts}
\makeatletter

\newcommand{\Rmnum}[1]{\expandafter\@slowromancap\romannumeral#1@}
\makeatother

\begin{document}

\title{Prolate-to-oblate transition and backbending along the
yrast line induced by quasiparticle alignment}

\author{Fang-Qi Chen}\email{fangqi0591@nwpu.edu.cn}
\affiliation{School of Physical Science and Technology, Northwestern
Polytechnical University, Xi'an 710129, China}

\author{Q. B. Chen}
\affiliation{Physik-Department, Technische Universit\"{a}t M\"{u}nchen,
D-85747 Garching, Germany}

\date{\today}
%%%%%%%%%%%%%%%%%%%%%%%%%%%%%%%%%%%%%%%%%%%%%%%%%%%%%%%%%%%%%%%%%%%%%%%%%%%%%
%%%%%%%%%%%%%%%%%%%%%%%%%%%%%%%%%%%%%%%%%%%%%%%%%%%%%%%%%%%%%%%%%%%

\begin{abstract}

The yrast lines in Kr isotopes with $N=42$, 44, and 46 are investigated in a
beyond mean field framework with both prolate-oblate coexistence and
quasiparticle alignment taken into account. Quasiparticle orbitals
with high-$j$ and low-$\Omega$ on the oblate side are shown to be
responsible for the sharp backbending observed in $^{82}$Kr, by
driving the yrast shape from prolate to oblate. This suggests that
quasiparticle alignment may not be neglected in the investigation
of the shape evolution along the yrast line.

\end{abstract}

%\pacs{21.10.Re, 21.60.Jz, 23.20.Js, 27.70.+q}% PACS, the Physics and Astronomy

\maketitle

The Kr isotopic chain from the neutron-deficient to the neutron-rich
side shows a rich collection of structure phenomena and has
attracted experimental and theoretical interests over the years.
Nuclear shape in the ground and low-lying states is one of the
topics discussed intensively in this aspect. For example, coexisting
$0^+$ states with different deformations~\cite{Piercey1982PRC,
Chandler1997PRC,Bouchez2003PRL} and rapid shape transitions with increasing
neutron numbers~\cite{Gade2005PRL} and so on
has been observed and understood in terms of various mean-field~\cite{Xiang2012NPA,
Abusara2017PRC} and beyond-mean-field~\cite{Bender2006PRC, Girod2009PLB,
Fu2013PRC, Yao2014PRC, Rodriguez2014PRC} theories. In these theoretical
studies it is noted that beyond-mean-field effects like symmetry restoration
or mixing of different shapes can be important in understanding the
remarkable phenomena observed~\cite{Fu2013PRC}.

Besides that along the isotopic chain, the shape evolution along the
yrast line is another interesting aspect in the structure studies. In
this case it demands a proper understanding for the high-spin region,
in which the pair breaking due to the Coriolis effect, i.e., the
quasiparticle alignment, may play an essential role. For example,
lifetime measurements in $^{74}$Kr have found pronounced reduction of
prolate deformation after the $S$-band crossing~\cite{Algora2000PRC},
which is described by the total Routhian surface (TRS) calculation.

The TRS calculation~\cite{Satula1994NPA} which provides energy
surfaces at specified rotational frequencies, is a powerful tool in
revealing the shape evolution along the yrast line. However, as a
mean field description the rotational symmetry remains broken, and
the configuration mixing is not allowed in the TRS calculation. On
the other hand, the projected shell model (PSM)~\cite{PSMreview, Sun2016PS}
is another powerful framework for the high-spin spectra, with the
rotational symmetry restored and the configuration mixing taken into
account. The PSM has been applied to high-spin studies of various deformed
mass regions, including the Kr isotopes~\cite{Palit2001NPA, Sun2001PRC,
Verma2006EPJA, Liu2015SC, Wu2017NPA}. One of the greatest success of
the PSM is the description of the backbending phenomenon induced by
the quasiparticle alignment.

The backbending described by the PSM involves the breaking of pairs
formed by quasiparticle orbitals with high-$j$ and low-$\Omega$ that
are sensitive to the Coriolis effect. These aligned orbitals are
defined at the deformation that is adopted for the ground state.
This identity of deformations of the vacuum and the multi-quasiparticle
configurations results from the assumption of a fixed deformation
parameter in the PSM, which is good for the well deformed
nuclei, but not necessary for the transitional nuclei such as
the Kr isotopes. Moreover, the fixed deformation assumed in the
PSM leaves little room for the shape coexistence in the ground
state, as well as the shape evolution with increasing spin.

As a generalization of the PSM, various deformations are taken into
account in the framework proposed in Refs.~\cite{FQChen2016PRC, FQChen2017PRC},
in which HFB vacua of different deformations, as well as two-quasiparticle
configurations based on each of them, are involved in the configuration space.
Such a framework can be expected to provide a proper description to the
transitional nuclei, in which the shape coexistence or shape
evolution might exhibit.

In particular, with the allowance of various shapes a new manner of
backbending might be anticipated. This kind of backbending involves
the breaking of oblate pairs when the ground state is basically prolate.
While the energy of the oblate vacuum increases quickly with spin,
the corresponding aligned two-quasiparticle configuration can dive
into the yrast region due to the strong Coriolis effect. Therefore,
the backbending in this case takes place simultaneously with a
prolate-to-oblate shape transition. Similar to the typical backbending
described in traditional PSM calculations, the backbending here
involves also high-$j$ and low-$\Omega$ orbitals, which locate near
the top of a high-$j$ sub-shell in the oblate case. Therefore,
this kind of backbending might happen when the Fermi level come
close to the top of a high-$j$ sub-shell. This is in contrast to
the typical backbending in the well deformed case, in which a Fermi
level near the bottom of a high-$j$ subshell is expected.
The existence of such backbending phenomenon originates from
the opposite orders of members in a high-$j$ shell with prolate and
oblate deformations. Note that the same character of Nilsson levels
also leads to the existence of oblate high-$K$ isomers related to
the $\pi h_{11/2}$ shell as reported very recently~\cite{Lizarazo2020PRL}.

In this paper the yrast lines of Kr isotopes with $N=42$, 44, and 46
are studied within the framework proposed in Refs.~\cite{FQChen2016PRC,
FQChen2017PRC}, in order to show the effect of oblate pair breaking on
the yrast behavior.

The model framework adopted in this work has been presented in
Ref.~\cite{FQChen2016PRC} and only a brief outline is given here.
The Hamiltonian used in the model is of the pairing plus quadrupole
type~\cite{PSMreview}
\begin{equation}\label{Hamiltonian}
 \hat{H}=\hat{H}_0-\frac{\chi}{2}\sum_\mu\hat{Q}^+_\mu\hat{Q}_\mu
 -G_M\hat{P}^+\hat{P}-G_Q\sum_\mu\hat{P}^+_\mu\hat{P}_\mu,
\end{equation}
which is diagonalized in a model space consists of a series of
axially symmetric HFB vacua:
\begin{equation}\label{vacua}
 |\Phi(\beta)\rangle,
\end{equation}
as well as two-quasiparticle states built on them:
\begin{equation}\label{2QP}
 \alpha^+_i(\beta)\alpha^+_j(\beta)|\Phi(\beta)\rangle.
\end{equation}
The HFB vacua are determined by the variation
\begin{equation}
 \delta\langle\Phi(\beta)|\hat{H}-\lambda_n\hat{N}
 -\lambda_p\hat{Z}-\lambda_q\hat{Q}_0|\Phi(\beta)\rangle=0,
\end{equation}
with the constraint on particle numbers and quadrupole moments,
while the quasiparticle operators $\alpha_i(\beta)$ are defined as
\begin{equation}
 \alpha_i(\beta)|\Phi(\beta)\rangle=0.
\end{equation}
Both of the vacua (\ref{vacua}) and the two-quasiparticle
configurations (\ref{2QP}) are projected onto good angular momentum
and particle numbers, so the ansatz for the wave function reads
\begin{equation}\label{wavefunction}
 |\Psi^I_\sigma\rangle=\sum_\rho f^I_{\sigma\rho}\hat{P}^I_{MK}
  \hat{P}^N\hat{P}^Z|\Phi_\rho\rangle,
\end{equation}
with the index $\rho$ running over the deformation $\beta$ as well as
the vacua and two-quasiparticle configurations $\{i,j\}$ considered. The
coefficients $f^I_\rho$ are determined by solving the Hill-Wheeler
equation
\begin{equation}
 \sum_{\rho'}[\mathcal{H}^I(\rho,\rho')
  -E^{I}_\sigma\mathcal{N}^I(\rho,\rho')]f^I_{\sigma\rho'}=0,
\end{equation}
with the norm and energy overlaps defined as
\begin{equation}\label{overlaps}
\begin{split}
 \mathcal{N}^I(\rho,\rho')&\equiv\langle\Phi_\rho|\hat{P}^I_{KK'}
  \hat{P}^N\hat{P}^Z|\Phi_{\rho'}\rangle,\\
 \mathcal{H}^I(\rho,\rho')&\equiv\langle\Phi_\rho|\hat{H}
  \hat{P}^I_{KK'}\hat{P}^N\hat{P}^Z|\Phi_{\rho'}\rangle.
\end{split}
\end{equation}
Note that the projected basis
\begin{equation}
 \{\hat{P}^I_{MK}\hat{P}^N\hat{P}^Z|\Phi_\rho\rangle\}
\end{equation}
is not orthogonal normalized~\cite{PSMreview} so the coefficients
$f^I_{\sigma\rho}$ can not be understood as the probability
amplitude for the intrinsic configuration $|\Phi_\rho\rangle$.
The probability amplitudes are defined as~\cite{ManyBody}
\begin{equation}
 g^I_\sigma(\rho)=\sum_{\rho'}\Pi^{1/2}(\rho,\rho')f^I_{\sigma\rho'}
\end{equation}
with $\Pi^{1/2}(\rho,\rho')$ obtained by the diagonalization of
the norm matrix (\ref{overlaps}):
\begin{align}
 \sum_{\rho'}\mathcal{N}^I_{\rho\rho'}u^I_{\kappa\rho'}
  &=n^I_\kappa u^I_{\kappa\rho},\\
 \Pi^{1/2}(\rho,\rho')
  &=\sum_\kappa u^I_{\kappa\rho}\sqrt{n^I_\kappa}u^{I*}_{\kappa\rho'}.
\end{align}
Therefore, $|g^I_\sigma(\rho)|^2$ are understood as the weight of
the configuration $|\Phi_\rho\rangle$ in the wave function
(\ref{wavefunction}) and satisfy the normalization condition as
\begin{equation}
 \sum_\rho|g^I_\sigma(\rho)|^2=1.
\end{equation}

In our calculations the neutron and proton major shells with
$N=2$, 3, and 4 are considered as the single-particle model space.
The spherical single-particle energies for the single-particle part
$\hat{H}_0$ in (\ref{Hamiltonian}) are determined by the Nilsson
scheme at $\beta=0$, with a modification for the Nilsson parameters
$\kappa$ and $\mu$ taken from Ref.~\cite{Nilsson1985}. The $\kappa$
and $\mu$ for the neutron shell $N=4$ are multiplied by a factor
1.12 and those for the same proton shell are multiplied by 1.3.
These modifications are done in order to obtain reasonable energy
curves as noted in Ref.~\cite{FQChen2016PRC}. The strengths $\chi$
of the quadrupole-quadrupole interactions are $90\%$ of those in the
earlier PSM calculation for $^{78}$Kr~\cite{Liu2015SC} and are kept
constant for all of the isotopes studied. The monopole pairing
strengths $G_M$ also come from the earlier PSM calculations~\cite{Palit2001NPA,
Liu2015SC}, and the quadrupole pairing strength are $G_Q=0.16G_M$.
The energy cutoff for the intrinsic configurations $|\Phi_\rho\rangle$
are the same as that in Ref.~\cite{FQChen2016PRC}. The effective
charges in the $B(E2)$ calculations are $e^\pi=1.35e$ for proton and
$e^\nu=0.35e$ for neutron, adjusted to get an overall agreement
with the observed transition strength $B(E2,2^+_1\rightarrow 0^+_1)$.

\begin{figure}[htbp]
\centering
\includegraphics[width=7.0 cm]{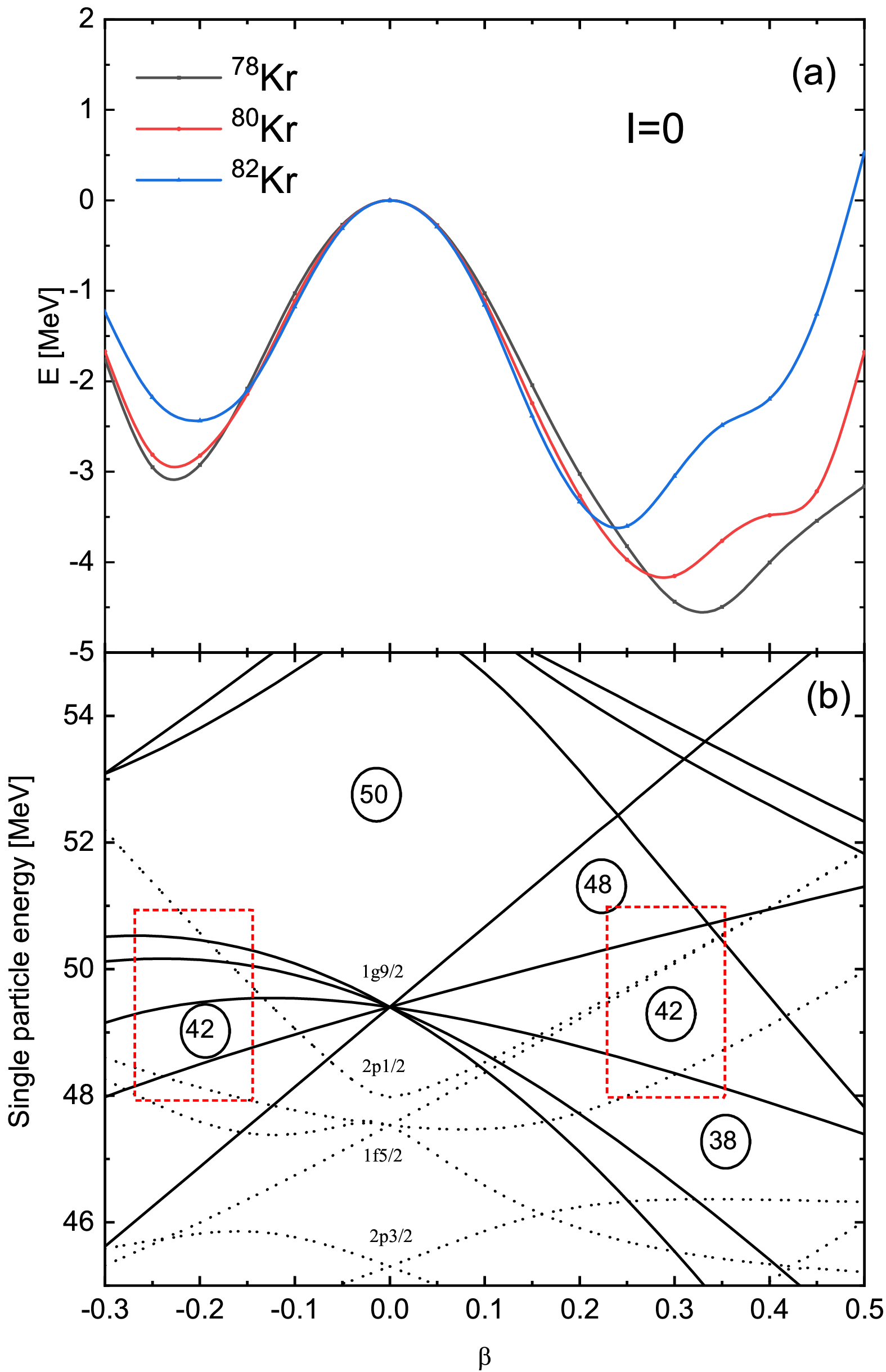}
\caption{(a) Projected energy curves at $I=0$ for $^{78-82}$Kr.
         The energies are normalized to $\beta=0$ in each curve. (b) Deformed neutron single-particle levels, calculated for $^{82}$Kr
         as an example. The solid (dashed) lines denote positive (negative)
         parity states. The Fermi level at the prolate/oblate minimum locates
         in the area circled by the dotted rectangles.}
\label{Fig12combined}
\end{figure}

As a preparation for the following discussions, an overall impression
of the mean field aspects of the Kr isotopes studied are
given in Fig.~\ref{Fig12combined}.
Fig.~\ref{Fig12combined}(a) shows the projected energy curves at $I=0$.
Two minima with prolate and oblate deformations can be found for
all of the three isotopes. The prolate minimum in $^{78}$Kr locates
at $\beta \sim 0.33$, in coincidence with the deformation
parameter adopted in the earlier PSM calculation~\cite{Liu2015SC}.
The oblate minimum lies about 1.5 MeV above the prolate one,
suggesting the prolate dominance in the ground and low-spin states.
The deformation of the prolate (absolute) minimum tends
to decrease with the neutron number, in accordance with the
increasing excitation energy observed for the $2^+_1$ states~\cite{NNDC}.
It suggests that the neutron Fermi level locates in the
upper half of the $\nu g_{9/2}$ sub-shell, moving towards its
top as the neutron number increases. In this case, the active $g_{9/2}$
orbitals around the Fermi level are the high (low)-$\Omega$ ones
for the prolate (oblate) side, as can be inferred from the Nilsson
diagram for neutron shown in Fig.~\ref{Fig12combined}(b). Note that the
location of the Fermi level and the prolate deformed ground state
fulfill the requirements for a favored oblate quasiparticle alignment,
as discussed above. Moreover, Fig.~\ref{Fig12combined}(a) show that
the energy difference between the two minima gets lower
as the neutron number increases, suggesting a growing importance
of the oblate configurations.

\begin{figure}[htbp]
\centering
\includegraphics[width=7.0 cm]{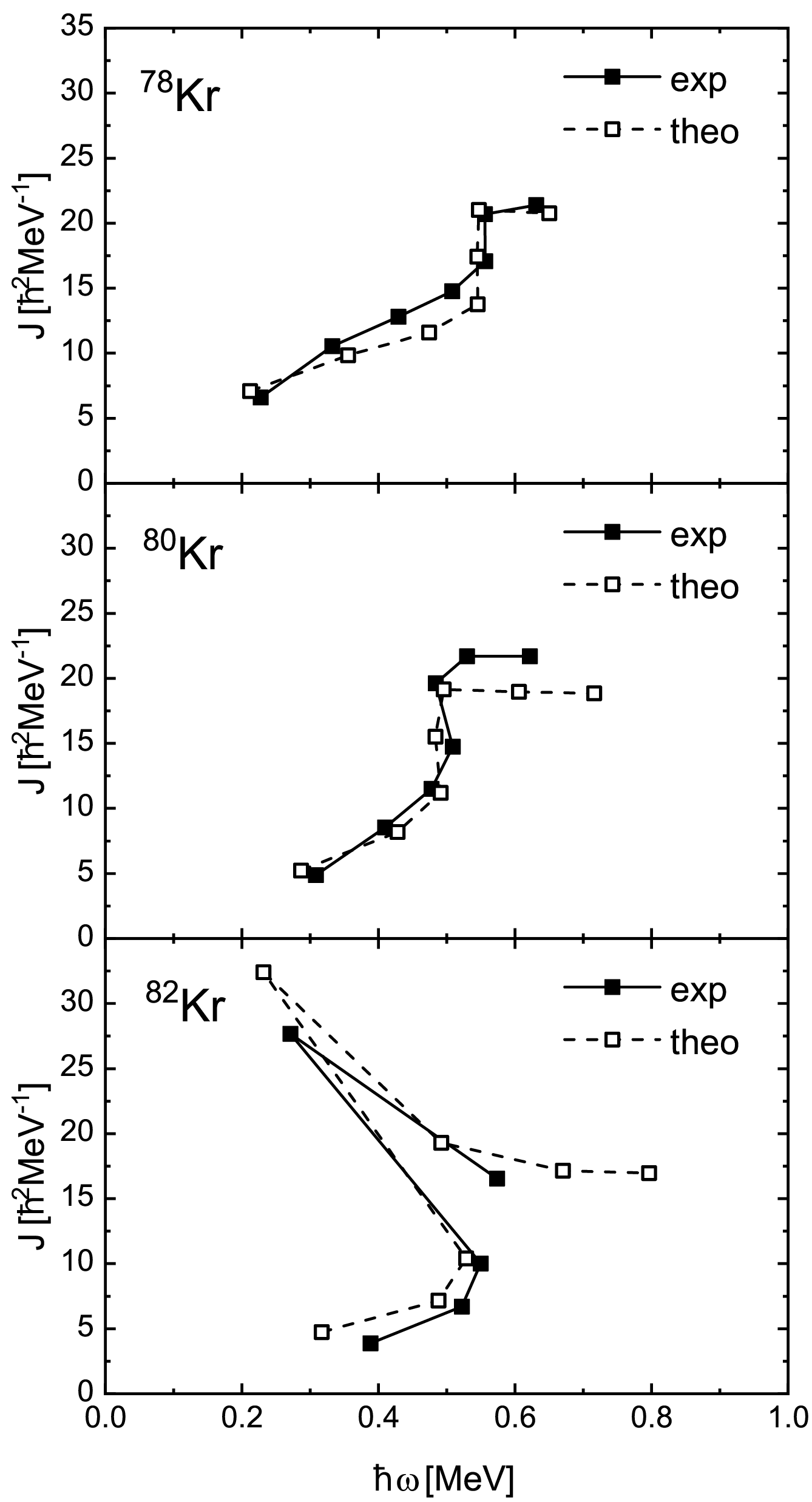}
\caption{The calculated moment of inertia as functions of the rotational
         frequency for the yrast bands in $^{78-82}$Kr, in comparisons with
         the experimental data taken from Ref.~\cite{NNDC}.}
\label{MoI-omega}
\end{figure}

The obtained yrast spectra are compared with the experimental ones~\cite{NNDC}
in Fig.~\ref{MoI-omega} in terms of the so-called ``backbending plot",
which gives the moment of inertia (MoI)
\begin{equation}
 \mathfrak{J}=\frac{2I-1}{E(I)-E(I-2)}
\end{equation}
as a function of the rotational frequency
\begin{equation}
 \hbar\omega=\frac{E(I)-E(I-2)}{2}.
\end{equation}
Such a plot is sensitive to the irregularities in the yrast band and
is widely used in the studies of band crossing phenomena.
The agreement between the experimental data and the calculated results
is good. The problem with the MoI at $I=2$ found in the earlier PSM
calculation~\cite{Liu2015SC} is now solved with the consideration of
the oblate part. As the neutron number increases, the irregularities
in the yrast bands take place earlier and become stronger, showing
an enhanced tendency of pair breaking. This is in contrast to the
shell filling pattern on the prolate side discussed above, as the
high-$\Omega$ $\nu g_{9/2}$ orbitals around the Fermi level are less
sensitive to the Coriolis effect. Therefore, the irregular behavior
in the yrast bands shown in Fig.~\ref{MoI-omega} indicates the possible
existence of oblate pair breaking.

\begin{figure}[htbp]
\centering
\includegraphics[width=6.0 cm]{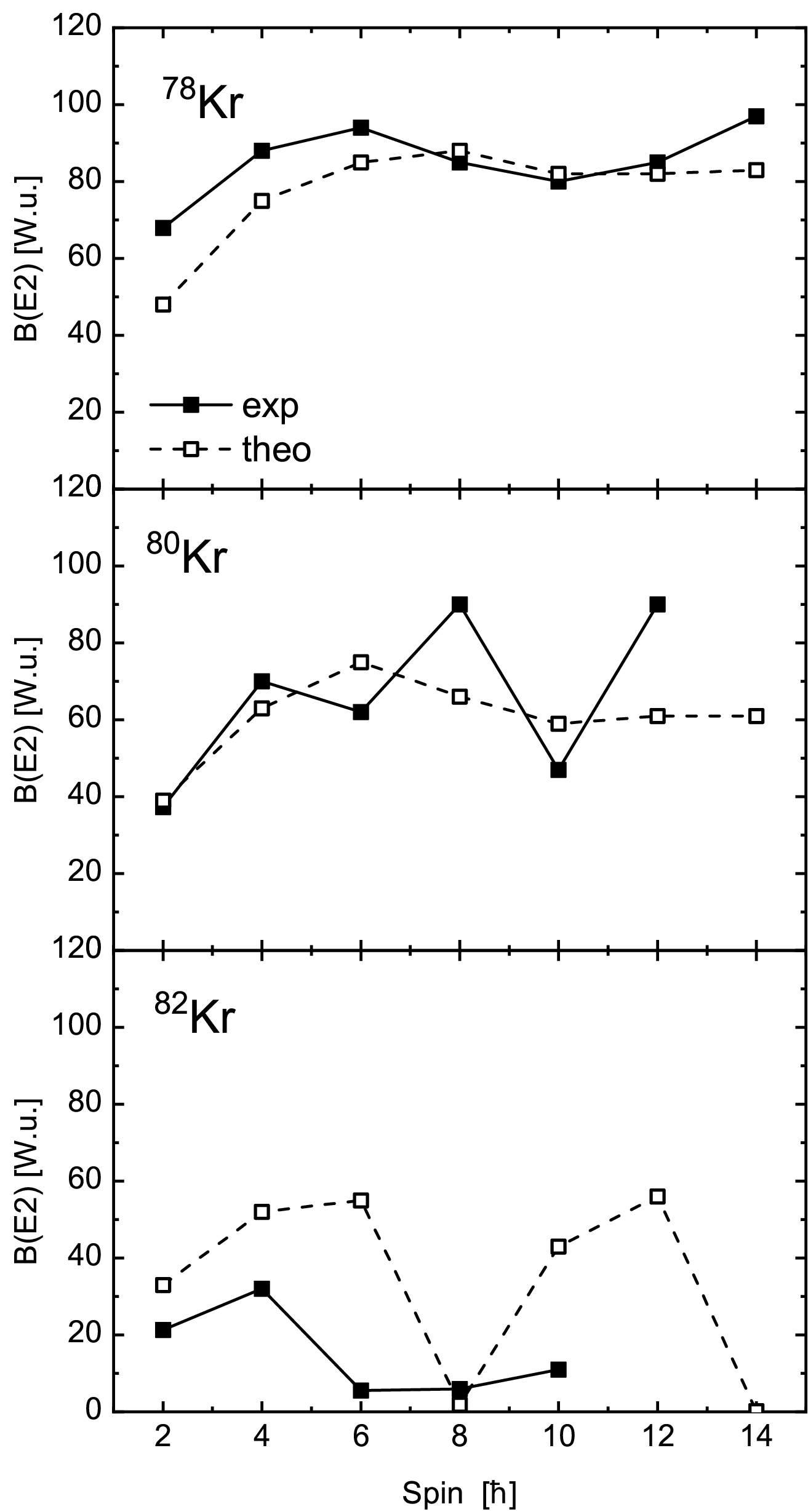}
\caption{The calculated $B(E2)$ along the yrast lines in $^{78-82}$Kr,
compared with the experimental data taken from Ref.~\cite{NNDC}.}
\label{B(E2)}
\end{figure}

As another important observable, the calculated strength of the $E2$
transitions along the yrast line are presented in Fig.~\ref{B(E2)},
in comparison with the experimental data~\cite{NNDC}. An overall
agreement between the theoretical and the experimental results is
achieved. For the low-spin part the decreasing trend of the $B(E2)$
versus the neutron number is qualitatively reproduced, in line
with the reduced tendency of deformation shown in Fig.~\ref{Fig12combined}(a).
In $^{78}$Kr a dip of the $B(E2)$ is found at around $I=10$, where the
irregularity in MoI emerges (see Fig.~\ref{MoI-omega}), as a reflection
of some structures change taking place. Such kind of dips get more
pronounced in $^{80}$Kr and $^{82}$Kr, in accordance with the enhanced
MoI irregularity, indicating abrupt structure change in these isotopes.
In particular, the dip found in $^{82}$Kr is so pronounced that the
$B(E2)$ at $I=8$ is of only the order of 1 W.u.. Such a remarkable
dip, together with the dramatic backbending shown in Fig.~\ref{MoI-omega},
suggests little interaction between the two crossing configurations,
which are likely to have, therefore, very different structures.

\begin{figure}[htbp]
\centering
\includegraphics[width=8.0 cm]{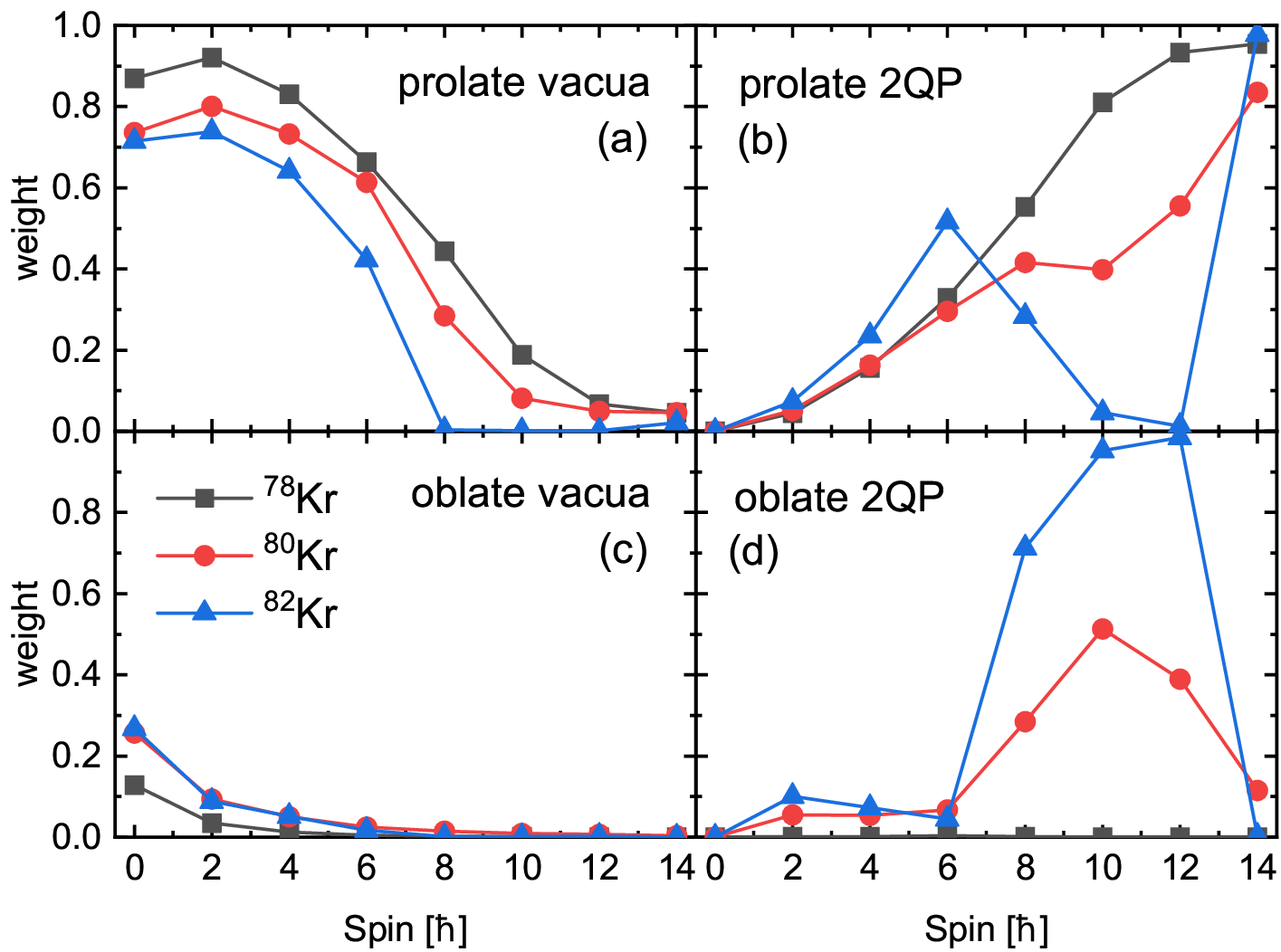}
\caption{Weight of various types of configurations calculated for the
         yrast bands in $^{78-82}$Kr: (a) prolate vacua; (b) prolate
         two-quasiparticle configurations; (c) oblate vacua; (d) oblate
         two-quasiparticle configurations.}
\label{weight2}
\end{figure}

The above discussions on the yrast spectra and $E2$ transitions
hint the appearance of oblate pair breaking and the consequent
crossing between prolate and oblate configurations. The actual
existence of these mechanisms can be illustrated in Fig.~\ref{weight2},
in which the weight $|g_{\sigma}^I(\rho)|^2$ of various types of
configurations are presented as functions of spin. The total
weight of the prolate deformed vacua are given in panel
Fig.~\ref{weight2}(a) for the three isotopes. For all of them the
weight of the prolate vacua are close to 1 at low spins, in accordance
with the prolate dominance shown in the energy curves (see Fig.~\ref{Fig12combined}(a)).
The largest value is found in $^{78}$Kr, who has the largest energy
difference between the prolate and oblate minima. The weight
of the prolate vacua decrease with spin, as a result of the band
crossing. Note that the slope of the decreasing reaches maximum
at the spin range where the irregularity in MoI takes place, due to
the relatively rapid configuration change in the crossing region.
The contributions from the prolate vacua become negligible after the
band crossing.

The weight of the prolate two-quasiparticle configurations are given
in Fig.~\ref{weight2}(b). They are small at the beginning and
increase smoothly before the band crossing. In this stage
the prolate dominance remains in the yrast states and the
increased weight here are contributed by the alignment of the
prolate quasiparticles. As discussed above, this kind of alignment
is moderate and does not lead to dramatic backbending in the MoI. In
the band crossing region the behavior of the three isotopes become
very different. For $^{78}$Kr the weight of the prolate
two-quasiparticle configurations keeps increasing until the maximum
value close to 1. This suggests that the irregularity (upbending) in
the MoI of $^{78}$Kr can be related to the crossing of two prolate
configurations (the vacuum and the two-quasiparticle configuration
from $\nu g_{9/2}$ orbitals), in consistent with the band crossing
pattern provided in traditional PSM calculations. The weight of the
prolate two-quasiparticle configurations for $^{80}$Kr stop
increasing at the point of band crossing and keeps smaller than that
for $^{78}$Kr afterwards. This behavior is inconsistent with the
stronger MoI irregularity observed in $^{80}$Kr, indicating
ingredients other than the prolate quasiparticle alignment may play
a role in this case. Finally for $^{82}$Kr, the weight of the
prolate two-quasiparticle configurations experience dramatic
decrease in the crossing region. Having little contribution from the
prolate part, the yrast states right after the crossing point must
be basically oblate. The sharp backbending observed in $^{82}$Kr is
therefore induced by the crossing between the prolate and oblate
configurations, which is likely to be abrupt due to the small
overlap between the two. Furthermore, it seems that in $^{82}$Kr
the prolate two-quasiparticle configurations become dominant again
at $I=14$, being consistent with the second drop of the theoretical
$B(E2)$ shown in Fig.~\ref{B(E2)}. However, this is the result
obtained without the consideration of four-quasiparticle
configurations, which may play a role at such a high spin.
Therefore, the interpretation of the $I=14$ member of the yrast
band remains open and more experimental information is needed.

As the counterpart of panel Fig.~\ref{weight2}(b), the weight of the
oblate two-quasiparticle configurations are shown in Fig.~\ref{weight2}(d).
The large bump at the crossing region shown for $^{82}$Kr confirms
the oblate nature of the yrast states in the range $I=8$-$12$,
supporting the prolate-to-oblate transition proposed above.
A smaller bump is found for $^{80}$Kr in the same spin range,
showing the oblate contribution, coexisting with the prolate part,
that is responsible for the upbending behavior. For $^{78}$Kr the
oblate two-quasiparticle configurations makes no contribution, in
consistent with the traditional band crossing pattern suggested.

In contrast to the oblate two-quasiparticle configurations, the
oblate vacua make little contribution to the yrast bands in the
whole spin range investigated, as shown in Fig.~\ref{weight2}(c).
The weight around 0.2 at $I=0$ drops quickly with spin and becomes
almost vanish before the crossing region. This is in accordance
with the well known fact that the rotated oblate minima is unfavored
compared with the prolate ones, as can be seen in Fig.~8(a) of
Ref.~\cite{Fu2013PRC}, for instance.

\begin{figure}[htbp]
\centering
\includegraphics[width=6.0 cm]{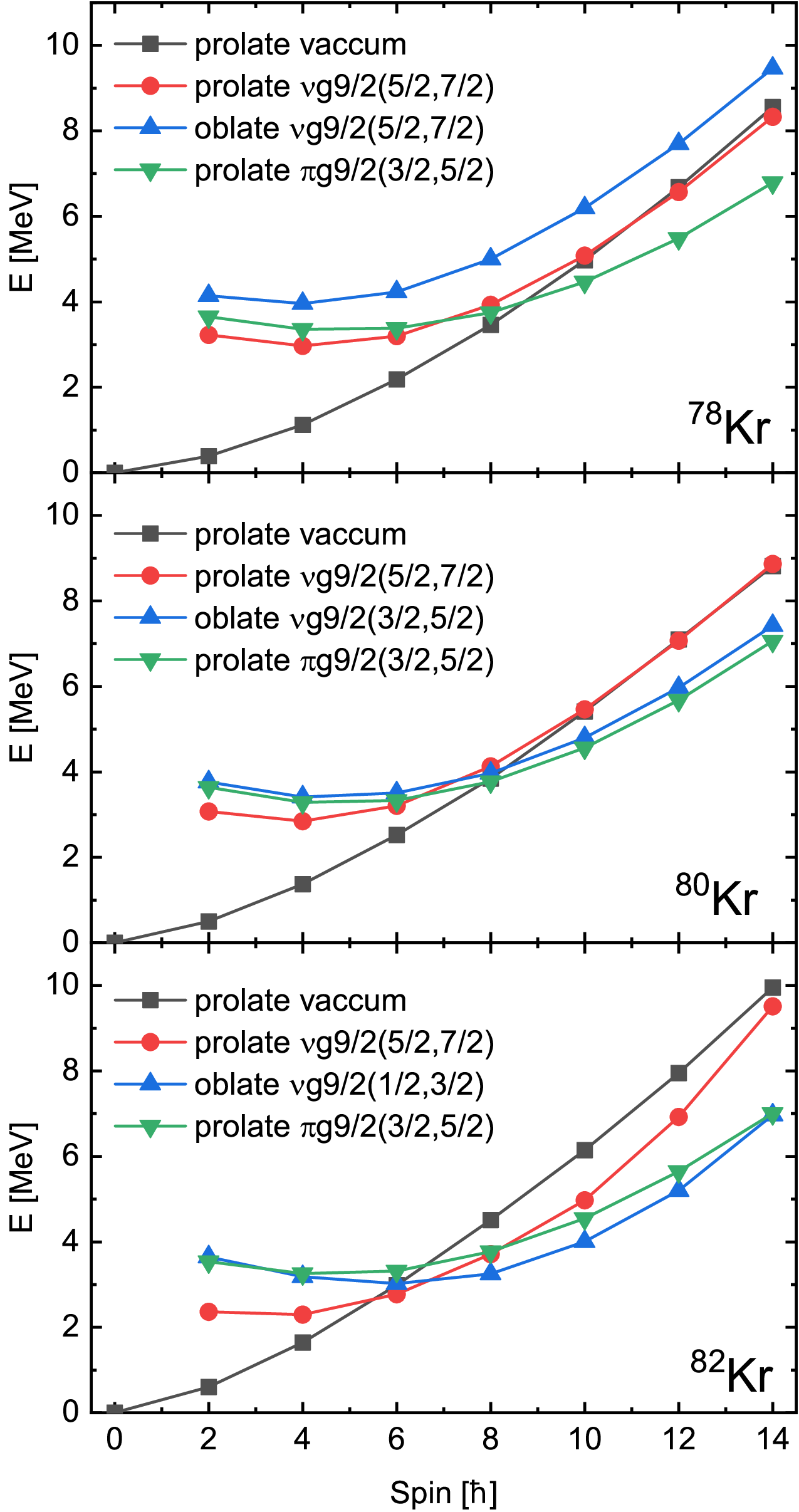}
\caption{The band diagrams defined as a generalization of those in PSM (see text)
         for $^{78-82}$Kr, with those configurations relevant in the yrast region
         included.}
\label{banddiagrams}
\end{figure}

For a better understanding of the yrast line behavior, quasiparticle
orbitals that are relevant to the crossing configurations need to be
specified, as in the discussions on the traditional backbending
phenomena. For this purpose one refers to the so-called ``band diagram",
which is widely used in the PSM analysis~\cite{PSMreview}. A ``band"
in such a diagram is formed by projecting a specific intrinsic
configuration (with the deformation fixed in the PSM) onto various spins,
and the configuration corresponding to the lowest band is recognized
as the dominant one for the yrast state. However, with the variation
of deformation opened up in our framework, the concept of the band diagram
has to be generalized. For example, the two-quasiparticle configuration
$\nu g_{9/2}(5/2,7/2)$ (here 5/2 and 7/2 are the quantum numbers of the projection
of the single particle angular momentum onto the three-axis of the intrinsic
frame) can be built on a series of prolate vacua, with deformations $\beta=0.20$,
$0.25$, $0.30$, etc. Each of them can be projected, forming a subspace at
each given spin in which the Hamiltonian can be diagonalized. The lowest
energy obtained in such a diagonalization corresponds to the optimal
combination of the above deformations, under the given configuration
$\nu g_{9/2}(5/2,7/2)$, and is thus defined as the ``band energy" in
the present case.

The band diagrams described above are presented in Fig.~\ref{banddiagrams}
for $^{78-82}$Kr, with those relevant configurations close to the Fermi
surfaces taken into account. The variation of the band crossing pattern
from $^{78}$Kr to $^{82}$Kr can be found in this plot. For $^{78}$Kr the
prolate $g$-band (projected from vacuum) is crossed by the prolate
two-quasiparticle configuration $\pi g_{9/2}(3/2,5/2)$, which is the
$\pi g_{9/2}$ configuration lying closest to the Fermi level. The lowest
$\nu g_{9/2}$ band on the prolate side, i.e., $\nu g_{9/2}(5/2,7/2)$, also
cross the prolate $g$-band but at a higher energy. On the other hand, the
lowest oblate band with the configuration $\nu g_{9/2}(5/2,7/2)$ appears at a
relatively high energy, away from the yrast region in the spin range investigated.
This is in accordance with the PSM-like band crossing pattern
obtained in the discussions of Fig.~\ref{weight2}. For $^{80}$Kr the
configuration of the lowest oblate band becomes $\nu g_{9/2}(3/2,5/2)$.
Unlike the oblate band in $^{78}$Kr, it crosses the prolate $g$-band
at around $I=8$, almost simultaneously with the prolate $S$-band
$\pi g_{9/2}(3/2,5/2)$. This is again in agreement with the conclusion
drawn from Fig.~\ref{weight2} that the upbending in $^{80}$Kr is
induced by both prolate and oblate two-quasiparticle configurations.
For $^{82}$Kr, the oblate band with the configuration $\nu g_{9/2}(1/2,3/2)$
crosses the prolate $g$-band at around $I=6$ and becomes the lowest one
afterwards, in accordance with the oblate-dominated band crossing suggested
in Fig.~\ref{weight2}. In general, a growing importance of the oblate
quasiparticle alignments in the band crossing region can be found from
$^{78}$Kr to $^{82}$Kr, in both the band diagrams and the weight calculations.

Such a growing importance of the oblate quasiparticle alignments can
be understood from two aspects: (1) the position of the Fermi level;
(2) the energy difference between the prolate and oblate minima.
As the neutron number increases, the neutron Fermi level moves
toward the top of the $\nu g_{9/2}$ sub-shell. With oblate deformations
the quantum number $\Omega$ of the orbitals around the Fermi level
become smaller in this process, as can be seen from the variation of
the oblate configurations adopted in the band diagrams for the three
isotopes. Therefore, the response to the Coriolis effect of the
oblate configurations gets enhanced and their band energies at
higher spins are thus suppressed. Moreover, the energy difference
between the two minima decrease from $^{78}$Kr to $^{82}$Kr, as
shown in Fig.~\ref{Fig12combined}(a). This makes the oblate configurations
energetically more favored as the neutron number increases and thus
more likely to play a role in the yrast region. Both aspects contribute
to the variation of the band crossing pattern shown in the band diagrams.

In summary, the yrast lines in Kr isotopes with $N=42$, 44, and 46
are investigated in this work. The model space consists of axially
deformed HFB vacua of both prolate and oblate shapes, as well as
explicit two-quasiparticle configurations based on them. Such a
model space accounts simultaneously for the shape mixture ranging from
the prolate to oblate and the pair breaking induced by Coriolis
interaction. The sharp backbending observed in $^{82}$Kr is interpreted
by the crossing between the prolate dominated vacuum and an oblate
two-quasiparticle state, involving high-$j$ and low-$\Omega$ orbitals
and is thus sensitive to the Coriolis effect. The oblate
quasiparticle alignment is also found to play a role in the
upbending observed in $^{80}$Kr, while in $^{78}$Kr its effect
is negligible. The variation of the importance of the oblate
two-quasiparticle states in the band crossing region in these Kr
isotopes is understood in terms of the location of the neutron Fermi
levels as well as the energy difference between the prolate and oblate
minima. The band crossing pattern revealed here shows that a
prolate-to-oblate shape transition can be induced by the quasiparticle
alignment, accompanied by the backbending, and it is thus suggested
that two-quasiparticle configurations might be necessary in the
investigations on the shape evolution along the yrast line.

\begin{acknowledgments}

Inspiring and valuable suggestions from L. J. Wang are acknowledged. F. Q. Chen thanks her mother in law, G. X. Guo, for taking care of
her baby during the quarantine due to COVID-19, which allows her to
finish this work. This work is partly supported by the National Natural
Science Foundation for Young Scientists of China under Grant No. 11905172,
the Fundamental Research Funds for the Central Universities under
Grants No. G2019KY05105, and the Deutsche Forschungsgemeinschaft
(DFG) and National Natural Science Foundation of China (NSFC)
through funds provided by the Sino-German CRC 110 ``Symmetries and
the Emergence of Structure in QCD'' (DFG Grant No. TRR110 and NSFC
Grant No. 11621131001).

\end{acknowledgments}

%%%%%%%%%%%%%%%%%%%%%%%%%%%%%%%%%%%%%%%%%%%%%%%%%%%%%%%%%%%%%%%%%%%%
%%%%%%%%%%%%%%%%%%%%%%%%%%%%%%%%%%%%%%%%%%%%%%%%%%%%%%%%%%%%%%%%%%%%%%%%%%%%%%%%%%%%%%%

\end{document}